\documentclass[usenatbib]{mnras}
\usepackage{newtxtext,newtxmath}
\usepackage{graphicx}
\usepackage{amsmath,amsfonts,amscd} 
\usepackage{microtype}
\usepackage{float}
\usepackage{verbatim}
\usepackage[T1]{fontenc}
\usepackage{ae,aecompl}
\usepackage{graphicx}

\usepackage{deluxetable}

\title[Physics of saturation in reconnection]{Physics of the saturation of particle acceleration in relativistic magnetic reconnection}
\author[D. Kagan, E. Nakar, and T. Piran]{Daniel Kagan $^{1}$\thanks{E-mail:kagandan@post.bgu.ac.il},  Ehud Nakar$^{2}$,Tsvi Piran$^{3}$\\
${^1}${Physics Department, Ben-Gurion University of the Negev, Be'er-Sheva 84105, Israel}\\
${^2}${Raymond and Beverly Sackler School of Physics \& Astronomy,  Tel Aviv University, Tel Aviv 69978, Israel}\\
${^3}${Racah Institute of Physics, The Hebrew University, Jerusalem 91904, Israel}}

\begin{document}
	\label{firstpage}
	\pagerange{\pageref{firstpage}--\pageref{lastpage}}
	\maketitle

\begin{abstract} We investigate the saturation of particle acceleration in relativistic reconnection using two-dimensional particle-in-cell simulations at various magnetizations $\sigma$. We find that the particle energy spectrum produced in reconnection quickly saturates as a hard power law that cuts off at $\gamma\approx4\sigma$, confirming previous work. Using particle tracing, we find that particle acceleration by the reconnection electric field in X-points determines the shape of the particle energy spectrum. By analyzing the current sheet structure, we show that physical cause of saturation is the spontaneous formation of secondary magnetic islands that can disrupt particle acceleration. By comparing the size of acceleration regions to the typical distance between disruptive islands, we show that the maximum Lorentz factor produced in reconnection is $\gamma \approx 5\sigma$, which is very close to what we find in our particle energy spectra.  We also show that the dynamic range in Lorentz factor of the power law spectrum in reconnection is $\le 40$. The hardness of the power law combined with its narrow dynamic range implies that relativistic reconnection is capable of producing the hard narrowband flares observed in the Crab Nebula but has difficulty producing the softer broadband prompt GRB emission.
\end{abstract}


\section{Introduction}
Magnetic reconnection is a process in which topology change in the magnetic field structure results in the rapid conversion of magnetic energy into kinetic energy. In the nonrelativistic case, this process has been directly observed on the Sun \citep{schmieder_15} and in the Earth's magnetosphere \citep{milan_17}. In the relativistic reconnection regime (see \citet{kagan_15} for a review), where the ratio of the magnetic energy to the total enthalpy of the particles (the magnetization $\sigma$) is much larger than 1, efficient particle acceleration in reconnection has been used to explain high-energy nonthermal emission from pulsar wind nebulae (PWN) \citep{kirk_03,  sironi_11,petri_12,uzdensky_11,cerutti_12b,cerutti_13a}, active galactic nucleus (AGN) jets \citep{2009MNRAS.395L..29G,giannios_13,nalewajko_11,narayan_12,petropoulou_16}, and the prompt phase of gamma-ray bursts (GRBs) \citep{thompson_94, lyutikov_03,giannios_05,lyutikov_06,icmart_11, mckinney_12,zhang_14,beniamini_13, beniamini_16,beniamini_grad_17}. 

Particle-in-cell (PIC) simulations are the most common method for probing the acceleration of particles in plasmas, including acceleration in relativistic reconnection.  In general, the physics of reconnection as revealed by PIC simulations is similar in two and three dimensions \citep{sironi_14, guo_14, werner_17a} unless the initial reconnection structure is inherently three-dimensional. This justifies the use of two-dimensional simulations which reduce computational cost and increase the system sizes that can be simulated.  These simulations generally focus on the pair plasma case which is both realistic and easy to simulate. In the limit for which all species are relativistic after undergoing reconnection, these pair plasma simulations capture the physics of electron-ion plasmas as well \citep{guo_16a, werner_16b}. The PIC simulations have shown that relativistic reconnection can produce power law energy spectra of the form $dN/d\gamma \propto \gamma^{-p}$ and accelerate particles efficiently to high energies \citep{2001ApJ...562L..63Z,zenitani_05b,zenitani_07, zenitani_hesse_08b,jaroschek_08b,bessho_05,bessho_07,bessho_12,daughton_07,lyubarsky_liverts_08, liu_11, cerutti_12b, cerutti_13a, cerutti_14, werner_16, sironi_14,sironi_16,guo_14,guo_15,guo_16b,liu_15,yuan_16,lyutikov_16,kagan_16}.

 While these results indicate that reconnection produces efficient particle acceleration, there are several difficulties in understanding how it occurs.  First, it is unclear what is the dominant acceleration mechanism in reconnection.  The simplest and most fundamental acceleration mechanism is linear acceleration by the reconnection electric field. However, other possible mechanisms also exist, many of which were reviewed by \citet{oka_10}. In various studies,  "anti-reconnection" between colliding islands \citep{oka_10,sironi_14}, curvature drift acceleration \citep{guo_14, guo_15}, and island contraction \citep{drake_06} were found to dominate the particle acceleration process in reconnection.

 The energy spectra produced by particle acceleration in reconnection also have several puzzling properties. The power law index is not universal but varies significantly with the magnetization $\sigma$ from $p>2$ at low $\sigma$ to $p\sim 1$ at high $\sigma$\citep{sironi_14,werner_16, guo_14,guo_15, lyutikov_16,kagan_16}. In contrast, PIC simulations of relativistic shocks show that a first-order Fermi process produces a power law spectrum with an approximately universal index of  $p\approx 2.2$ \citep{sironi_15rev}. This indicates that the power law produced in reconnection results not from a self-similar acceleration process but self-consistently from another physical constraint. 
 
An additional feature of the power law energy spectra in reconnection is that $1<p<2$ for large $\sigma$. As a result, the total energy in a logarithmic interval, proportional to $\gamma^2 dN/d\gamma$, is dominated by the highest-energy particles. This immediately indicates that this power law must have finite extent to satisfy energy conservation. Indeed, \citet{werner_16} have found that for large $\sigma$  a cutoff to the power law occurs at $\gamma_{f}\approx4 \sigma$ independent of $\sigma$. However, the variation in $p$ with $\sigma$ makes it uncertain how energy conservation constraints are satisfied by this cutoff.

This paper investigates these questions about the causes of particle acceleration in reconnection and the properties and saturation of the particle energy spectrum resulting from that acceleration. We carry out two-dimensional PIC simulations of pair plasmas at many values of $\sigma$. We then calculate the properties of the particle energy spectrum, confirming that saturation occurs. We trace particles to show that acceleration by the reconnection electric field in X-points is the dominant particle acceleration mechanism in these simulations. Combining these results, we suggest that the saturation is caused by the spontaneous formation of secondary  magnetic islands, at whose edges the magnetic field is strong enough to deflect the particles.  By analysing the structure of the current sheet during reconnection, we show that this hypothesis correctly predicts both the presence of saturation and the energy at which this saturation occurs.

Our paper is organized as follows. In Section \ref{sec:methods}, we discuss our simulation methods. In Section \ref{sec:results}, we present our results. In Section 4, we summarize the implications of our results for reconnection models of radiation from high-energy astrophysical systems. Finally, in Section \ref{sec:conclusions}, we discuss the conclusions of our research.

\section{Methodology} \label{sec:methods}

We use the particle-in-cell (PIC) method to carry out our simulations. This method evolves the discretized exact equations of electrodynamics (the mean-field Maxwell's Equations and the Lorentz Force Law), allowing us to precisely probe the full physics of particle acceleration, limited only by the accuracy of the discretization. In the PIC method discretization is carried out by using macroparticles to represent many individual particles and tracking fields only on the vertices of a grid. We implement the PIC simulations using the {\tt Tristan-MP} code  \citep{spitkovsky_structure_2008}, which uses current filtering to greatly reduce noise even at relatively small macroparticle densities.  
\

Our initial conditions are very similar to those in \citet{werner_16}. This allows us to verify their results and directly probe the physics responsible for their conclusions regarding the saturation of reconnection. The spatial domain is rectangular with $0\leq x<L_x$, $0\leq y<L_y$, and the boundary conditions are periodic in all directions.  The initial configuration contains two relativistic Harris current sheets \citep{harris62,kirk_03} without guide field at  $x=L_x/4$ and $x=3L_x/4$ with equal and opposite currents.
This configuration is susceptible to the growth of reconnection from thermal noise. Each current sheet has a magnetic field given by 

\begin{equation}
  {\mathbf B}=B_0  \tanh \left(\frac{x-x_0}{\Delta}\right) (\pm \hat{{\mathbf y}}),
\label{eq:harris_field}
\end{equation}
where $\Delta$ is the half-thickness of each current sheet and $x_0$ is its center. The sign in $\pm$ is positive for the vicinity of the  current sheet at $x_0=L_x/4$ and negative near the current sheet at $x_0=3 L_x/4$.

The density profile of the particles consists of a specially varying, drifting current population with maximum density $n_0$ centered at each current sheet, plus a background population of stationary particles of density $n_{\rm b}$:

\begin{equation}
n=n_0 \ {\rm sech}^2\left(\frac{x-x_0}{\Delta}\right) +n_{\rm b}. \label{eq:density_profile}
\end{equation}
We define densities including both species.

 Our primary goal in our simulations is the probe the saturation of particle acceleration as a function of $\sigma$, the magnetization in the background plasma

\begin{equation}\sigma\equiv \frac{B_0^2}{4\pi m n_{\rm b} c^2 h},\label{eq:sigmadef} \end{equation}
where $h=\langle\gamma\rangle +P/(m n c^2)$ is the average enthalpy of background particles.

We carry out simulations for $\sigma=$3, 10,  30, 100, 200, and 500, which gives us enough resolution to explore the dependence of parameters on $\sigma$. We name the simulations with ${\tt Sx}$, where $x$ is the value of $\sigma$ for the simulation.

The stationary populations of both species begin in relatively cold relativistic Maxwellians at temperature $T_{\rm b}=0.05 m c^2$. We always choose $n_0=16n_{\rm b}$. To ensure pressure balance across the current sheet, the temperature $T_0$ of the particles in the current sheet must be given by 

\begin{equation}n_0 T_0=n_{\rm b} T_{\rm b}\left(\frac{\sigma h}{T_b}\right)=\sigma n_{\rm b} h.\end{equation}

Since our background particles are cold, we find $T_0\approx\sigma n_{\rm b}/n_0$. Therefore, we initiate all particles in the drifting population at temperature $T_0=\sigma/16$.

\subsection{Length and time scales}\label{sec:params}

 The particle acceleration process occurs at a rate proportional to the magnetic field $B_0$. In order to compare results at different values of $\sigma$, it is advantageous to use simulation parameters for which the rate of particle acceleration is the same in each simulation. This can be done by normalizing all length scales to the product $\sigma  r_{\rm L}$, where $ r_{\rm L}=m c^2/(q B_0)$ is the Larmor radius of particles moving at mildly relativistic speed in the background plasma.  Our results will show that most aspects of the particle acceleration process do not vary with $\sigma$ when quantities are normalized in this way.

 We set our grid spacing as $ \sigma  r_{\rm L}/32$ for most values of $\sigma$ ( $\sigma  r_{\rm L}/64$ for the simulation with $\sigma=500$) and the width $\Delta$ of the current sheet to $\Delta=(5/32)\sigma  r_{\rm L}$ for all values of $\sigma$.   Simulations with larger $\Delta=(5/16)\sigma  r_{\rm L}$ indicate that our results do not depend on $\Delta$.  We find that our simulations are well resolved with respect to the size of these small scales, with no significant differences in particle acceleration physics even for grid spacings as small as $\sigma  r_{\rm L}/128$ (with the same relative current sheet sizes).

\citet{werner_16} found that saturation in relativistic reconnection required that simulations have current sheet lengths larger than $40 \sigma  r_{\rm L}$.  The overall length scale of our simulations is typically $(L_x, L_y) =(320 \sigma  r_{\rm L}, 320\sigma  r_{\rm L})$ (half as large for $\sigma=500$).  All runs are carried out for a time of at least $180 \sigma  r_{\rm L}/c$ ($100 \sigma r_{\rm L}/c$ for Simulation {\tt S500}), which is more than long enough for us to observe saturation. We carry out an additional simulation for $\sigma=100$ with twice the length and time scales to verify that long-term saturation is indeed occurring.

To ensure sufficient resolution in density to capture the physics of reconnection, we use a density of 8 macroparticles/cell/species throughout the plasma. Tests of the code \citep{kagan_16} show that the physics of reconnection and the evolution of the current sheet is similar for macroparticle densities up to 50 macroparticles/cell/species.

  \section{Results} \label{sec:results}

We find that the evolution of reconnection is similar to that found in other 2D simulations, including that of \citet{werner_16}. The tearing instability produces small-scale reconnection regions and magnetic islands, which merge with time. In the meantime, fast particle acceleration occurs. We will first find a phenomenological fit to the particle energy spectrum and then use it to constrain the characteristics of the saturation of particle acceleration in Section \ref{sec:fitting}. We then investigate the properties of particle acceleration in the current sheet using test particle simulations in Section \ref{sec:traces}. Finally, we analyze the properties of the current sheet and how they physically explain saturation in Section \ref{sec:cursheetcalc}.

\subsection{Fitting of Particle Spectra} \label{sec:fitting}

\begin{deluxetable}{cccccccccc}
\tabletypesize{\scriptsize}
  \tablewidth{0pt}
\tablecolumns{10}
  \tablecaption{Fits of simulations at saturation ($\tau\approx 84$) \label{tab:overview}}
  \tablehead{\colhead{{Run}\tablenotemark{a}} & \colhead{{$p$}} & \colhead{{$\gamma_i$}} & \colhead{ $\gamma_f$} & \colhead{$\zeta$\tablenotemark{b}} & \colhead{$f_{\rm acc}$\tablenotemark{b,c}} & \colhead{ $\langle \gamma\rangle_{\rm acc}$\tablenotemark{d}} & \colhead{$\langle \gamma\rangle_{\rm pl}$\tablenotemark{e}} & \colhead{$f_{\rm int}$\tablenotemark{f}}}
\startdata
{\tt S3}&{2.23}&1.8    &6.4    &    1.5    &    0.42    &    3.1    &3.1&0 \\
{\tt S10}&1.72&2.4    &34    &    2.3    &    0.74    &    7.4    &8.2&0.037\\
{\tt S30}&1.46&2.9    &132    &    2.4   &    0.89    &    18.4    &20.4&0.026\\
{\tt S100}&1.24    &10    &    434    &    2.1    &    0.91    &    80    &88    &0.053\\
{\tt S200}&1.19    &19    &    847    &    2.1    &    0.92    &    167    &179&0.071\\
{\tt S500}&1.16    &36    &    2140&    1.9    &    0.95    &    394    &427&0.045    
\enddata

\tablenotetext{a}{The number for each run gives the value of $\sigma$.}
\tablenotetext{b}{These parameters continue to evolve slowly after saturation}
\tablenotetext{c}{The fraction of the particle kinetic energy in accelerated particles (those with Lorentz factor above $\gamma_i$). }
\tablenotetext{d}{ The average Lorentz factor of accelerated particles.}
\tablenotetext{e}{The average Lorentz factor predicted for accelerated particles based on our power law fits and Equation (\ref{eq:powerlawpred}).}
\tablenotetext{f}{The fraction of the particle kinetic energy in intermediate particles (those in the interval $1.8<\gamma<\gamma_i$). }
\end{deluxetable}

\begin{figure}
\begin{center}

\includegraphics[width = 0.47\textwidth]{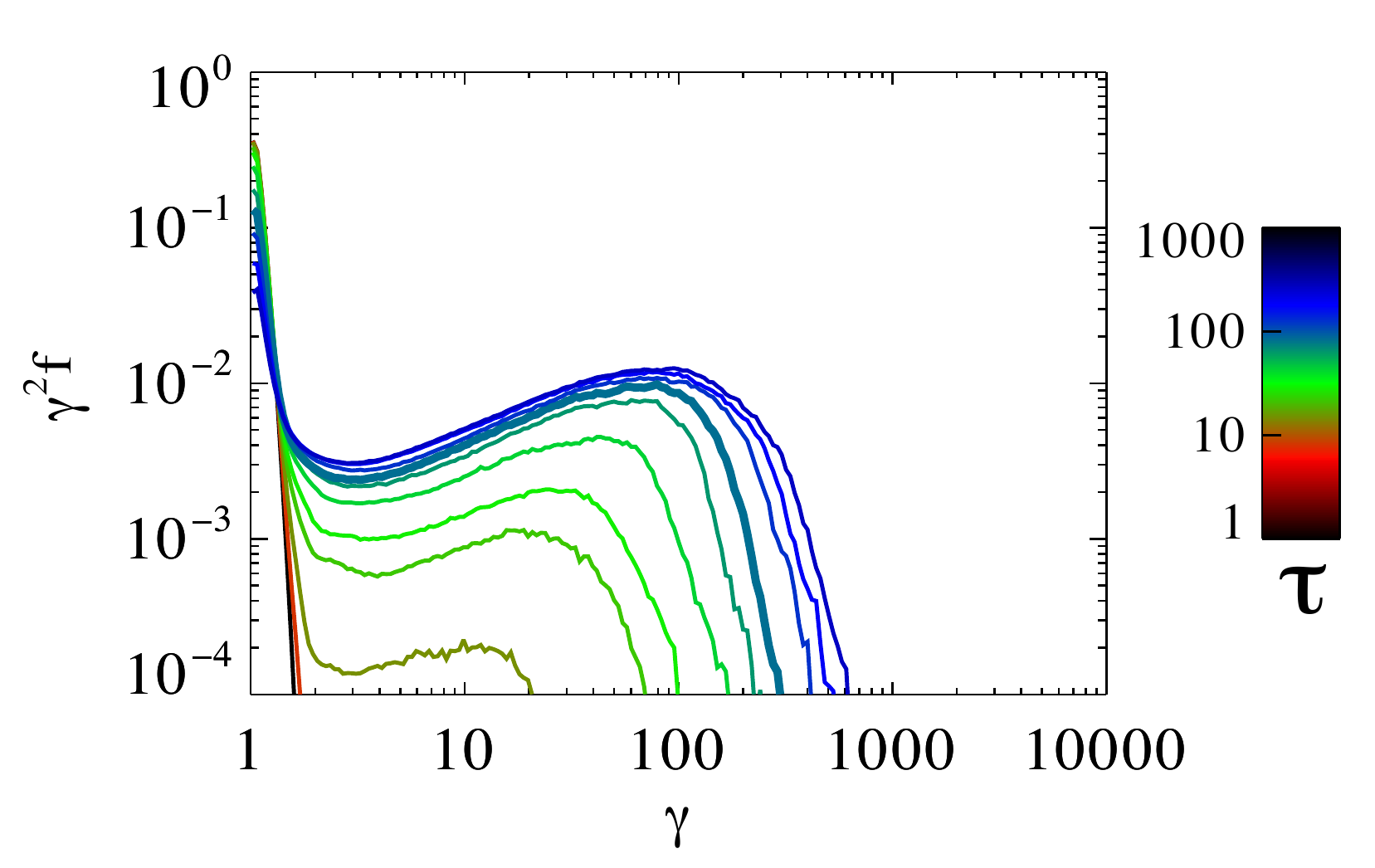}

\end{center}
\caption{ The evolution of the normalized particle energy spectrum $\gamma^2 dN/d\gamma=\gamma^2 f$ with the normalized time $\tau=c t/(\sigma  r_{\rm L})$ for Simulation {\tt S30}. Successive spectra are approximately logarithmically spaced. The spectrum at the time of saturation $\tau_{\rm s}\approx84$ is shown using a thicker line. Note that our energy spectra here and in other plots do not include the contribution of the hot population of particles that began the simulation in the current sheet.  
	\label{fig:sigma30evol}}
\end{figure}

Figure \ref{fig:sigma30evol} shows the evolution of the overall particle energy spectrum with normalized time $\tau=c t/(\sigma  r_{\rm L})$ for $\sigma=30$. Because successive spectra are logarithmically spaced, this plot provides clear evidence that saturation of the particle energy spectrum occurs near the time $\tau\approx100$. Our later analysis based on a phenomenological fit to the spectrum indicates that the time of saturation is approximately $\tau_{\rm s}\approx84$ for all values of $\sigma$.  This is in rough agreement with the results of  \citet{werner_16} that $L_{\rm min}=40 \sigma r_{\rm L}$  is the minimum length scale required for saturation. It corresponds to two light-crossing times for a current sheet of that length. 

\subsubsection{Saturation of the power law}

To constrain the properties of this saturation more precisely, we fit the high-energy spectrum  for simulations with all values of $\sigma$. 
Qualitatively, it is clear from the Figure \ref{fig:sigma30evol} that a thermal spectrum is present at low energies that fully cuts off at approximately $\gamma=1.8$. Then, an approximate power law begins at a location $\gamma_i$ and ends with a possibly exponential cutoff at location $\gamma_f$.

We make a phenomenological fit to the portion of the spectrum that begins at $\gamma_i$ (the start of the power law) with the function

 \begin{equation}
 N(\gamma)=A \left(\frac{\gamma}{\gamma_{i}}\right)^{-p} e^{-({\gamma/\gamma_{f})^{\zeta}}}.\label{eq:fitfunc} \end{equation}

This allows us to probe the properties of the power law and of the high-energy cutoff. The fitting of $\zeta$ allows us to compare our results with those of \citet{werner_16}, whose cutoff fitting function was the product of an exponential with $\zeta=1$ and one with $\zeta=2$. The value of $\gamma_i$ where we begin fitting must be input to the fitting function, so we do fits for various values of this parameter and compare them. We select values for $\gamma_i$ that are as small as possible without significantly worsening the fit.

\begin{figure}
\begin{center}
\includegraphics[width = 0.48\textwidth]{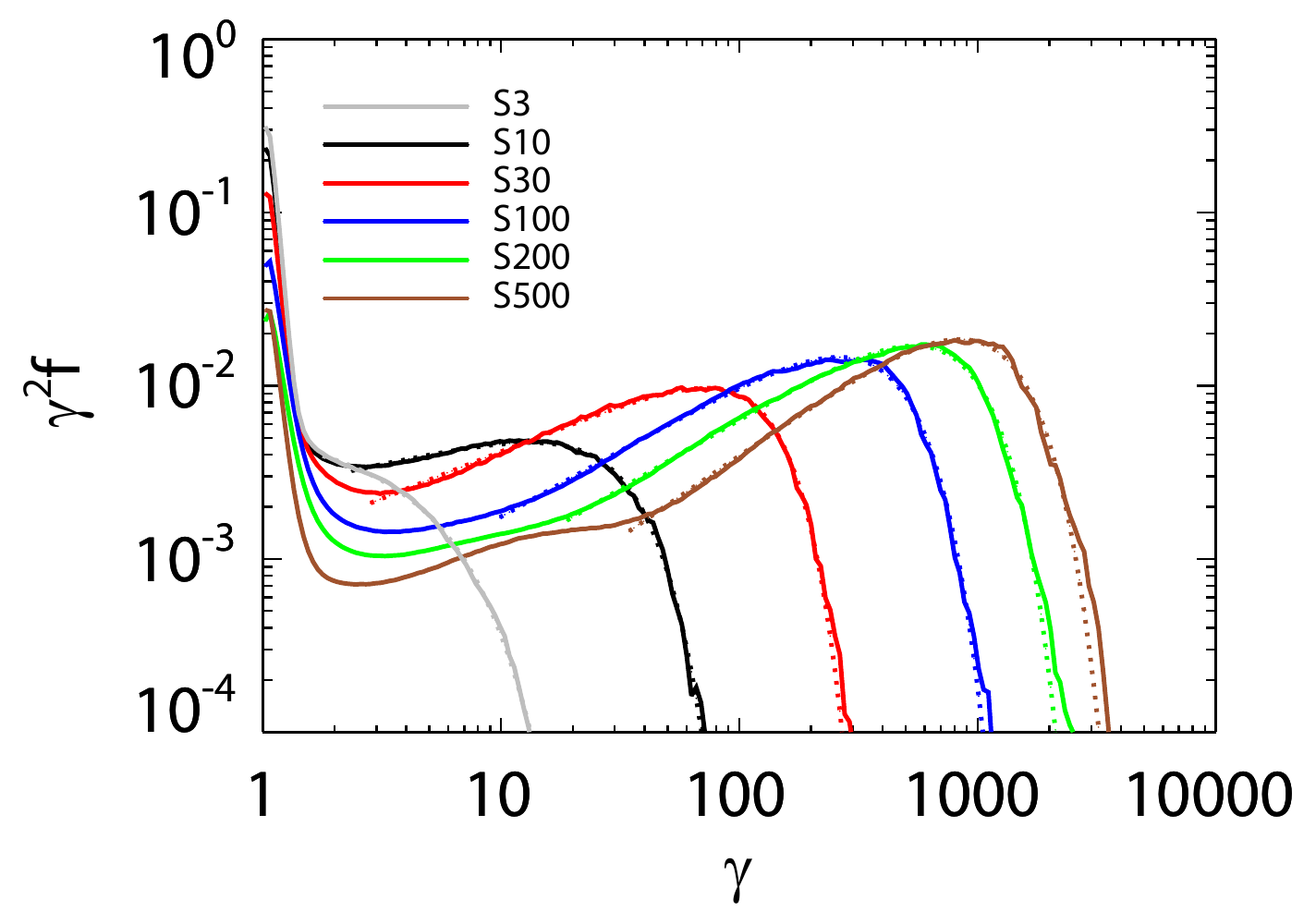}
\end{center}
\caption{ The normalized particle energy spectrum per logarithmic bin $\gamma^2 dN/d\gamma=\gamma^2 f$ at the time of saturation $\tau_{\rm s}\approx84$ for each value of $\sigma$. The calculated fits above $\gamma_i$ for each value of $\sigma$ are shown using dashed lines.\label{fig:fitcomp}}
\end{figure}

Our fits show that saturation occurs at approximately the same time for all values of $\sigma$, which is at normalized saturation time $\tau_{\rm s}\approx84$. The only fitting parameter that does not saturate at this time is $\zeta$, the shape of the high-energy cutoff. Table 1 shows the fitting parameters averaged over several inputs near this time of saturation, while Figure \ref{fig:fitcomp} visually shows the particle energy spectra (solid lines) and the fits (dotted lines) to those spectra.
The evolution of $p$ with $\sigma$  is generally consistent with that found by \citet{werner_16} both qualitatively and quantitatively. $p$ decreases monotonically with $\sigma$ to an approximate asymptote at $p\approx1.2$, with $p>2$ only for $\sigma=3$. We also confirm their findings with respect to $\gamma_f$, which is indeed approximately $4 \sigma$ for most values of $\sigma$. An exception to this is the case $\sigma=3$, for which in any case we do not have enough dynamic range to obtain a good power law fit.

The fraction $f_{\rm acc}$ of the particle energy in the power law at the saturation time $\tau_{\rm s}\approx84$ is also indicated in Table 1. The table indicates that the power law particle population already has $>90\%$ of the particle energy at saturation for $\sigma >10$. Even for Simulation {\tt S3}, $f_{\rm acc}>0.7$ at $2 \tau_{\rm s}$,  which  is a very short time on macroscopic scales. Thus, energy from the power law  population will dominate the radiation produced in reconnection, as generally assumed in models of this process.

Reconnection is an efficient process, as can be seen from the fact that the magnetization in the current sheet is of order unity (see Figure 3 of \citet{kagan_16} for an example showing this). We therefore expect that particles that experience full acceleration in the current sheet will receive an average energy of $\langle\gamma\rangle\approx \sigma$. Table 1 shows that $\langle \gamma\rangle\geq 0.6\sigma$ in all cases. 






 We now check how well our power law fits represent the overall energetics of the accelerated particles. For a power law with index $p$ beginning at $\gamma_i$ and ending at $\gamma_f$, the average particle energy is given by

\begin{equation}
\langle\gamma\rangle_{\rm pl}=\left(\frac{1-p}{2-p}\right)\frac{\gamma_{f}^{2-p}-\gamma_{i}^{2-p}}{\gamma_{f}^{1-p}-\gamma_{i}^{1-p}}. \label{eq:powerlawpred}\end{equation}

We note that the commonly used approximation in which the $\gamma_f$ term is dropped from the denominator and the $\gamma_i$ term from the numerator is highly inaccurate for $p<2$. Table 1 shows that the power law represents the overall energetics very well, within $\pm 10\%$, although it typically overestimates the average energy slightly.

We now discuss the uncertainties in our fits.  The fit has a significant dependence on the choice of error model, but the resulting variation is not overwhelming so long as the  fractional error in the distribution decreases as $N(\gamma)$ increases.  In the reported fits, we assume that errors in the spectrum are Gaussian in logarithmic space, proportional to  $1/\sqrt{\gamma N(\gamma)}$.   Uncertainties in the choice of error model and the choice of $\gamma_i$ dwarf the calculated error for most of the parameters, imposing a significant uncertainty of at least $\pm 0.1$ in the power law index $p$ and of $\pm 20\%$ in the location of the cutoff $\gamma_i$. The cutoff exponent $\zeta$ is very sensitive to the error model (but insensitive to $\gamma_i$), with approximate uncertainty $\pm 0.2$. The fitting uncertainties are significantly greater for the runs with $\sigma=3$ and $\sigma=500$. In the earlier case, the shortness of the power law portion of the spectrum makes it difficult to estimate the parameters of the fit, while in the latter case the shape of the spectrum appears to be less well described by our fitting function.  These uncertainties and the trend in their magnitudes are roughly consistent with those found by \citet{werner_16}.

\subsubsection{The lower limit of the power law and the intermediate population}
\label{sec:lowerlimit}

The value of $\gamma_i$, the lower limit of the power law, is not reported by \citet{werner_16}. But it is crucial to understanding the particle energy spectrum produced in relativistic reconnection. This is well illustrated by Figure \ref{fig:fitcomp}, which compares the fits of  particle spectra at saturation for all values of $\sigma$. While $\gamma_i \approx 2$ for low $\sigma$, it increases significantly at larger values of $\sigma$, reaching an approximate value of  $\sigma/10$ for Simulations {\tt S30}, {\tt S100}, and {\tt S200}. The low value of $\gamma_i=36\approx 500/13.5$ for Simulation {\tt 500} is not too surprising given the high uncertainty in the fit for this value of $\sigma$.

The coupled saturation of both $p$ and $\gamma_i/\sigma$ at large $\sigma$ is not a coincidence. Let us assume that $p$ is constant, and $\gamma_i =a \sigma$ and $\gamma_f=b \sigma$  (with $a$ and $b$ being constants). Then Equation (\ref{eq:powerlawpred}) becomes $\langle \gamma \rangle_{\rm pl}=  k \sigma $, where 

\begin{equation}k=\left(\frac{1-p}{2-p}\right)\frac{b^{2-p}-a^{2-p}}{b^{1-p}-a^{1-p}},\end{equation}
is a constant.  Thus, the coupled saturation of $p$ and $\gamma_i$ combined with the already present saturation of $\gamma_f $ ensures that the average energy of the power law is a constant proportional to $\sigma$, as expected for an efficient reconnection process.  

The fact that $\gamma_i$ is far above the end of the thermal distribution at $\gamma=1.8$   at high $\sigma$ implies the presence of a significant population of particles in the interval $1.8<\gamma<\gamma_i$. These particles are neither unaccelerated like the background plasma nor fully accelerated to $\langle\gamma\rangle \sim \sigma$. We do not detect any significant difference between the locations in the current sheet of the intermediate particles and of the highest energy particles. This indicates that the intermediate particles go through the current sheet but receive little energy during the reconnection process.

The intermediate part of the spectrum as shown in Figure 2 can be approximated as a power law with $p\sim 2$, although the shape of the spectrum changes with $\sigma$. Thus, it is possible that at very high $\sigma1000$ the intermediate component of the spectrum expands or disappears. Table 1 shows that for $\sigma>3$ these particles have $f_{\rm int}\approx 0.05$ of the particle kinetic energy at the time of saturation.  The variation in this number with $\sigma$ is not systematic, and may result from uncertainties in the value of $\gamma_i$. $f_{\rm int}$ does not change significantly following saturation, indicating that this part of the spectrum has also reached a steady state.  We leave the investigation of the detailed properties of these intermediate particles to future work.

The values for $\gamma_i$ shown in Table 1 imply that $\gamma_f/\gamma_i \approx40$ at large $\sigma$. This is a very narrow range and restricts the dynamic range of particle acceleration in reconnection. We discuss the implications of this result and the other properties of our fits for reconnection models of radiation from high-energy systems in  Section \ref{sec:astrophysobj}.

\subsubsection{Analysis of the high-energy cutoff}\label{sec:zetaevolution}

\begin{figure}
	
	\begin{center}
		
		\includegraphics[width = 0.48\textwidth]{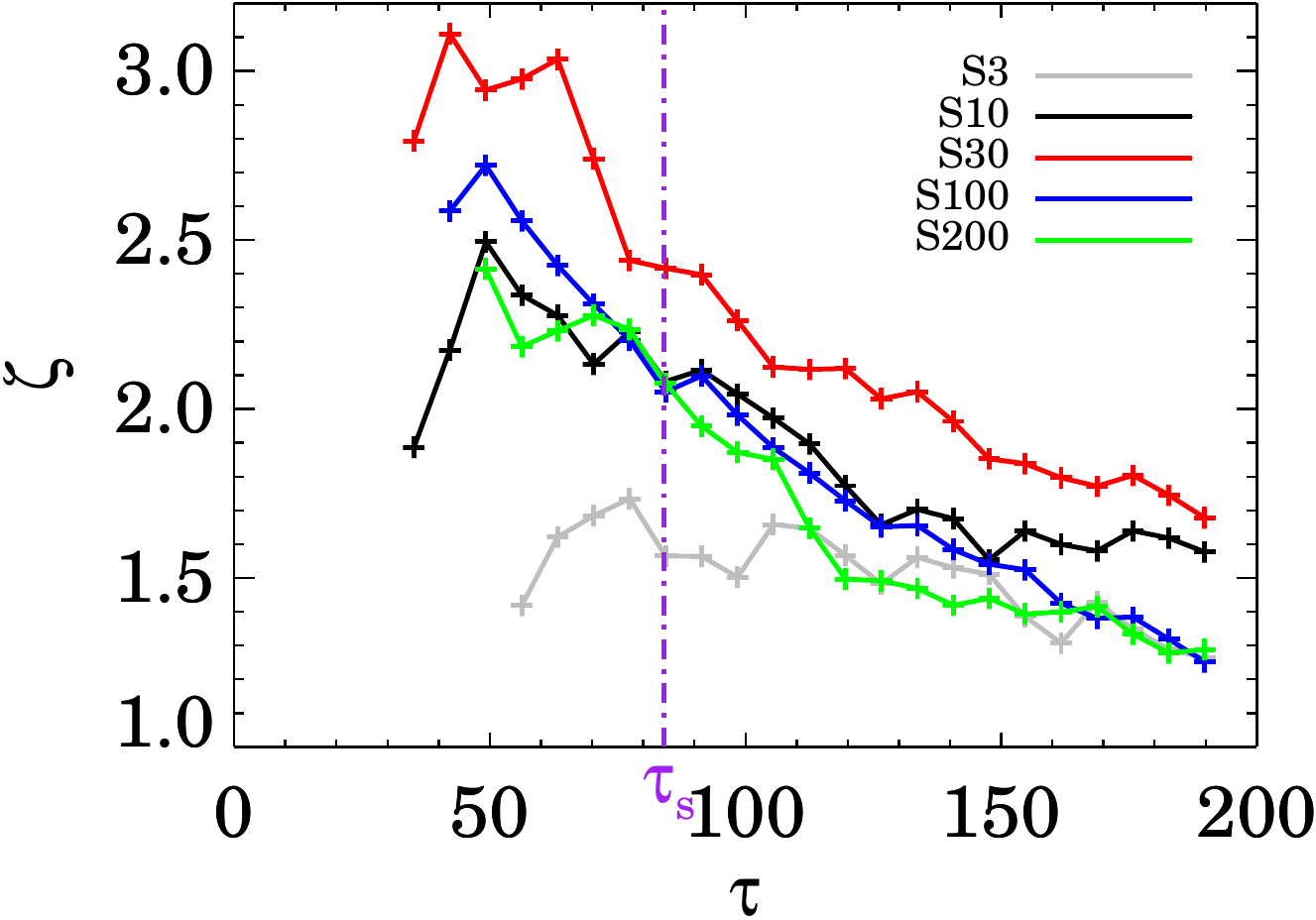}
	\end{center}
	\caption{ The evolution of $\zeta$ with normalized  time $\tau$ for all values of $\sigma$ except $\sigma=500$, for which the simulation was not run for a long enough time to probe its evolution well beyond the saturation time of the other parameters. Data points are shown only for times at which a good fit to Equation (\ref{eq:fitfunc}) could be found. The saturation time $\tau_{\rm s}\approx84$  is indicated using the dot-dashed line.  \label{fig:zetadependence}}
	
\end{figure}
Figure \ref{fig:zetadependence} shows how $\zeta$ depends on time for all values of $\sigma$ except $\sigma=500$.  At the time of saturation,  $\zeta \approx 2.4$ for Simulation {\tt S30} and $\zeta\approx 2.1$ for all other simulations with  $\sigma>3$. This is roughly consistent with the results of \citet{werner_16}, who found that at small system sizes the high-energy cutoff was well described with a cutoff corresponding to  $\zeta=2$. After $\tau_{\rm s}$, $\zeta$ decreases monotonically with time.  At $\tau=2 \tau_s$,  $\zeta\approx 1.3$ for Simulations {\tt S100} and {\tt S200}, $\zeta\approx1.6$ for Simulation {\tt S10}, and $\zeta \approx 1.7$ for Simulation {\tt S30}.

We have carried out some simulations at larger box size $L_x=L_y=500\sigma r_{\rm L}$ to  further investigate the evolution of $\zeta$.  They indicate that at late times the value of $\zeta$ continues to decrease, reaching approximately $\zeta\approx1.25$ for $\sigma=30$ and $\zeta\approx1.15$ for $\sigma=100$ at $\tau=4\tau_{\rm s}$.  Comparing these results with the evolution in Figure \ref{fig:zetadependence} indicates that the monotonic decrease in $ \zeta$ slows at late times, but we do not see a clear asymptote in this evolution. Our results are roughly consistent with a cutoff at very late times that is asymptotically a simple exponential $\zeta=1.0$, as suggested by the results of \citet{werner_16} for large reconnection systems. However, truly understanding the character of  the high-energy cutoff will require future simulations with enough computer resources to probe the evolution of the particle energy spectrum over temporal and spatial scales larger than our simulations by an order of magnitude.

\subsection{Analysis of particle acceleration}\label{sec:traces}

\begin{figure}

\begin{center}
\includegraphics[width = 0.48\textwidth]{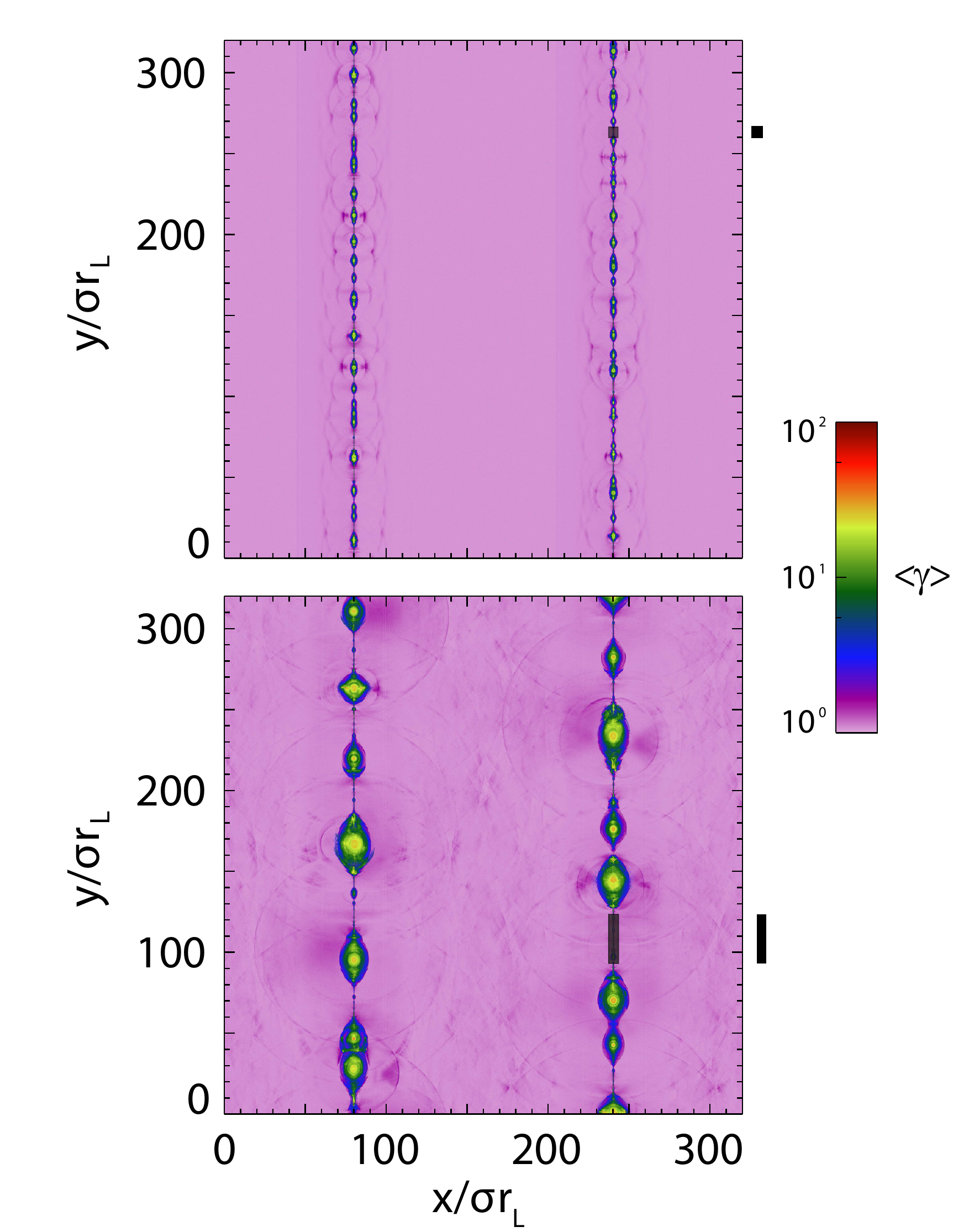}
\end{center}

\caption{The current sheet structure throughout the simulation with $\sigma=30$ displayed using the mean Lorentz factor $\langle \gamma \rangle$  at times $\tau\approx35$ before saturation (top) and at $ \tau\approx140$ after saturation (bottom). The shaded boxes show the locations of the reconnection regions centered at $(x_0,y_0)=(240,263.3) \sigma r_{\rm L}$ before saturation and $(x_0,y_0)=(240,110.75) \sigma r_{\rm L}$ after saturation where we use test particles to probe the physics of particle accelerations. The edges of the boxes show the locations of particle injection, which are at $x_0\pm 3\sigma  r_{\rm L}$ with a spread of  $2 \sigma r_{\rm L}$. We repeat these boxes to the right of the figure for clarity.\label{fig:cursheetstructure}}

\end{figure}

In order to find why the saturation of the power law at $\gamma_f \approx 4 \sigma$ is occurring in our simulations (and those of \citet{werner_16}), we must first identify the primary acceleration process in the simulations. There are two methods for assigning significance to acceleration mechanisms. The first is an arithmetic measure, which compares the difference in Lorentz factor $\gamma_{\rm after}-\gamma_{\rm before}$ before and after each acceleration process. The second is a logarithmic measure, which instead compares the ratios $\gamma_{\rm after}/\gamma_{\rm before}$ for each process. The latter is the more important measure because acceleration can occur over several orders of magnitude and thus the logarithmically dominant process determines the shape of the spectrum.

To investigate the primary acceleration processes before and after saturation, we trace particles entering acceleration regions in the simulation with $\sigma=30$ at $ c t/\sigma  r_{\rm L}\approx35$ before saturation and at $ c t/\sigma  r_{\rm L}\approx140$ after saturation. We initiate the particle tracing before and after saturation by duplicating particles flowing into X-points at these times and adding spread to their momenta and location.  We then trace the particles for a period of $ \tau\approx70\ (105)$ for the pre-saturation (post-saturation) tracing.

Figure \ref{fig:cursheetstructure} shows the structure of the simulation at these times, with the reconnection regions for which particle tracing is done highlighted with boxes and the initial locations of the traced particles entering the current sheet shaded in grey. From the figure, it is clear that the number of secondary islands in reconnection regions is increasing with time: there are no such islands in our chosen pre-saturation reconnection region, while there are between  2 and 9 islands of varying sizes in the post-saturation reconnection region (depending on the size needed for structures to be considered significant). Based on our more detailed study of current sheet structure in Section \ref{sec:cursheetcalc} we find that there are 3 significant islands in this X-point that can deflect high-energy particles above $\gamma=4\sigma$. The length of the primary reconnection regions has also increased approximately linearly with time, from $ D=6.25\sigma  r_{\rm L}$ at $ c t/\sigma  r_{\rm L}\approx35$ before saturation to  $ D=30\sigma  r_{\rm L}$ at $ c t/\sigma  r_{\rm L}\approx 140$ after saturation.

\begin{figure}

\begin{center}
\includegraphics[width = 0.4\textwidth]{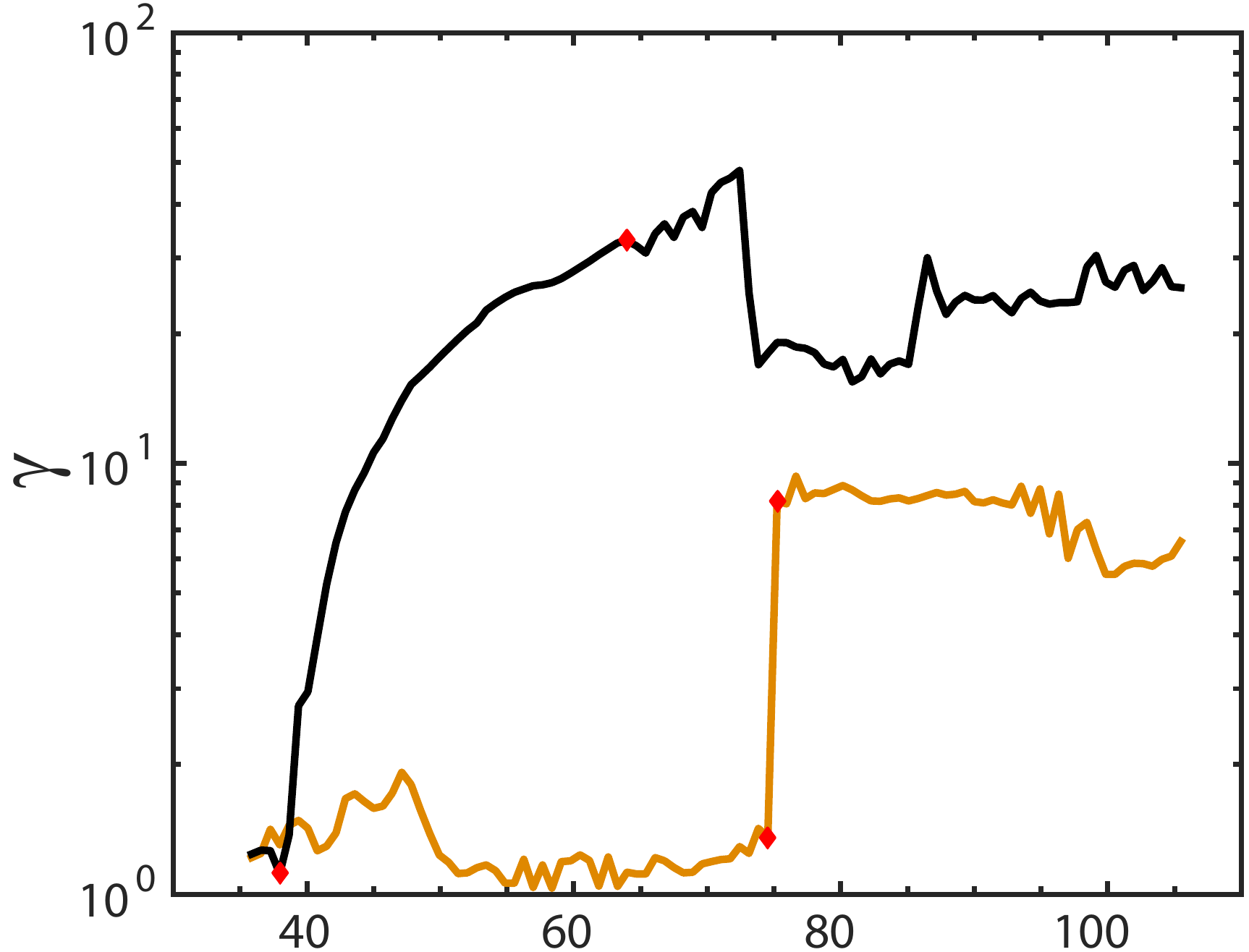}
\includegraphics[width = 0.4\textwidth]{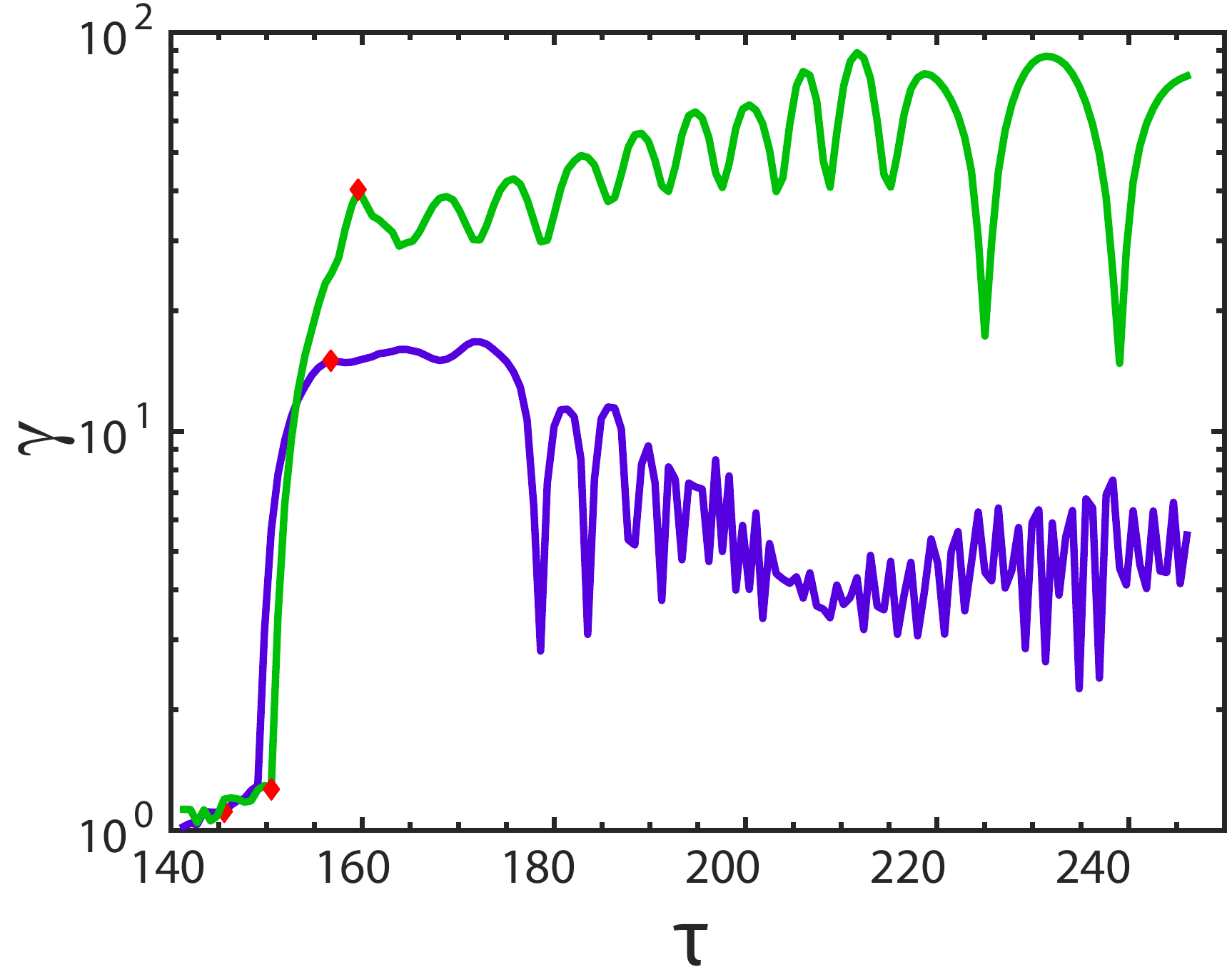}
\end{center}
\caption{The particle acceleration history for two typical particles in the pre-saturation particle tracing beginning at $\tau\approx35$ (top) and two other typical particles in the post-saturation particle tracing beginning at $\tau\approx140$ (bottom) in Simulation {\tt S30}. The particles are labeled with different colors as particles A (black), B (orange), C(green), and D (blue)  The majority of the acceleration for all particles clearly occurs in a single rapid acceleration episode, consistent with X-point acceleration. This rapid acceleration phase is bounded with red diamonds. \label{fig:acceltrace}}
\end{figure}

Figure \ref{fig:acceltrace} shows particle acceleration histories for two typical particles before and two other typical particles after saturation. It is clear from these figures that most of the energy gain for these particles occurs in a single, linear acceleration episode. For Particles C (green) and D (blue)  post-saturation and Particle A (black) pre-saturation, it is clear that this acceleration takes place in the shaded X-point where the particle was injected. Even Particle B (orange) in the pre-saturation run is accelerated in an X-point. The late time at which its acceleration begins is due to the fact that this X-point is not the one shaded in Figure \ref{fig:cursheetstructure} but the one just above it.

 We now consider the contribution of other acceleration mechanisms. Particle C (green) in the in the post-saturation run oscillates at late times as it reflects off of concentrations of $B_x$. This acceleration is clearly due to a Fermi process associated with either island contraction or curvature drift acceleration, as the acceleration per cycle increases with $\gamma$. However, this acceleration process increases the energy of the particle by a factor of only 2.

To confirm that these acceleration histories are indeed typical of the particle acceleration, we identify the rapid acceleration phase for each particle. This is done by selecting the period of monotonic increase in $\gamma$ during which it increases by the greatest factor.  Table 2 shows the averaged properties of the traced particles during the rapid acceleration phase for both the pre- and post- saturation tracing periods. We find that the particle energy at the end of this rapid acceleration phase is 13.7 (12.4) in the pre- (post-) saturation tracing. In  both cases, this is more than 2/3 of the average energy $\langle\gamma\rangle=18.4$ of all accelerated particles throughout the simulation with $\sigma=30$ shown in Table 1. This rapid acceleration also  dominates the total acceleration experienced by the traced particles in the X-point, taking up more than 90\% (60\%)  of the acceleration before (after) saturation. Furthermore, more than $80\%$ of particles in both traces begin the rapid acceleration phase of acceleration at $\gamma<2$, indicating that this phase is typically the first strong acceleration phase. This result favors X-point acceleration (independent of $\gamma$),  over Fermi-type acceleration in which the energy gain is proportional to $\gamma$.

 Logarithmically, X-point acceleration is even more dominant. X-point acceleration typically increases the particle energy by a factor of at least $12.4$, while all other mechanisms combined produce at most an increase by a factor of $\sim 18.4/12.4 \approx 1.48$. Thus, X-point acceleration clearly determines the shape of the spectrum in our simulations.

  \begin{deluxetable}{cccccc}
 \centering
  \tablewidth{0.45\textwidth}
  \tablecaption{Properties of the rapid acceleration phase \label{tab:overview}}

\tablehead{\colhead{\ $\tau_{\rm inj}$\tablenotemark{a}} &\colhead{\ $\langle\gamma\rangle_{\rm r}$\tablenotemark{b}\ } &$\langle\gamma\rangle_{\rm f}$&\colhead{\ $f_{<2}$\tablenotemark{d}\ }&\colhead{$\langle E_{z}\rangle_{ \rm r}/B_0$\tablenotemark{e}}&\colhead{\ $\langle v_y\rangle_{\rm r}/c$\tablenotemark{f}}}
\startdata 
35&13.7&14.1&0.88&0,25&0.50\\
140&12.4& 20.6&0.81&0.33&0.72
\enddata
\tablenotetext{a}{ The approximate normalized time of injection.}
\tablenotetext{b}{The average Lorentz factor at the end of rapid acceleration.}
\tablenotetext{c}{The average Lorentz factor at the end of X-point acceleration}
\tablenotetext{d}{The fraction of particles that begin rapid acceleration with $\gamma<2$.}
\tablenotetext{e}{The average electric field in the $z$ direction during the rapid acceleration phase, used in Section \ref{sec:cursheetcalc}.}
\tablenotetext{f}{The average velocity in the $y$ direction during the rapid acceleration phase, used in Section \ref{sec:cursheetcalc}. }
\end{deluxetable}

Our results indicate that X-point acceleration by the reconnection electric field is indeed the dominant particle acceleration mechanism, in both the arithmetic and especially the logarithmic sense. Many studies find that alternative mechanisms are important arithmetically \citep{sironi_14,guo_14,guo_15}.  But in the more important logarithmic sense, those studies usually\footnote[1]{In the study of \citet{guo_14} it is unclear whether X-point acceleration or a first-order Fermi process involving curvature drift is logarithmically dominant because some details of the acceleration process are not shown.  But later work from the same researchers \citep{guo_15} confirms that X-point acceleration is logarithmically dominant in their simulations.}  still find that X-point acceleration is dominant \citep{sironi_14, guo_15}. We note that this conclusion may change on extremely large scales not currently accessible to simulations, because acceleration during the many generations of island mergers that occur on such scales may become more energetically important than in current simulations. 

\subsection{Effects of current sheet structure on the particle acceleration process}\label{sec:cursheetcalc}

\subsubsection{Conditions for the disruption of particle acceleration}
In the previous section, we showed that the primary particle acceleration mechanism is acceleration by the electric field in the X-point. The energy equation for particles accelerated electromagnetically is given by
\begin{equation}
mc^2 \frac{d \gamma}{dt} = q \mathbf{E}\cdot \mathbf{v}.
\end{equation}

Particle acceleration in the X-point therefore requires that the particle's momentum is (anti)- aligned with the electric field in the reconnection region, but a particle can be deflected when it encounters a  magnetic island of sufficient size and magnetic field strength, disrupting the acceleration process. But since the deflection actually occurs at the edge of the island (or equivalently the adjacent X-point), we characterize both X-points and islands in a unified way, referring to them generically as current sheet structures (CSS).
\begin{figure}
	\begin{center}
		\includegraphics[width = 0.4\textwidth]{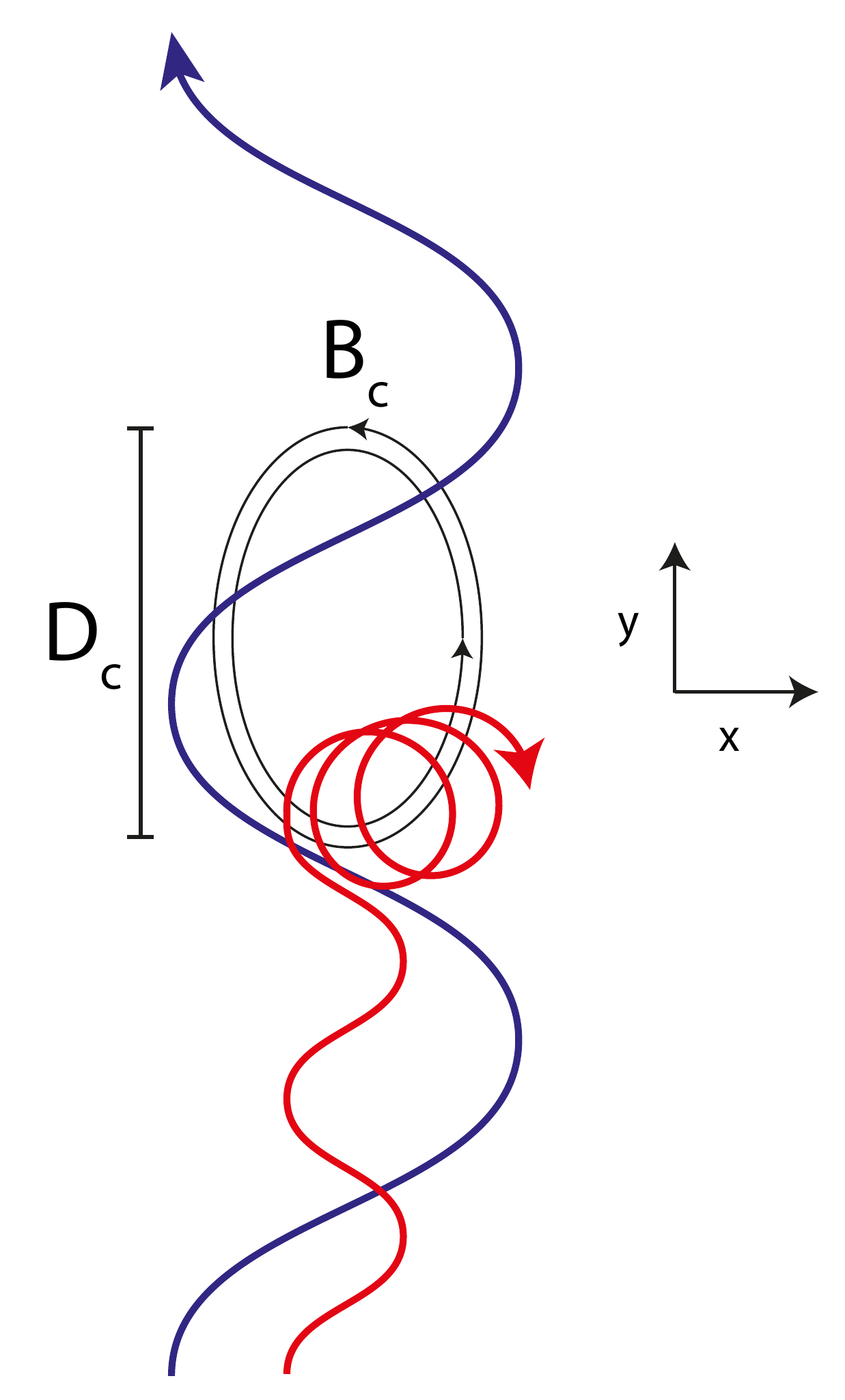}
	\end{center}
	\caption{ A schematic figure comparing the typical trajectories of a particle entering a magnetic island with $\gamma \ll \gamma_{c}$ (red) and with $\gamma\gg \gamma_{\rm c}$ (blue). The red particle is quickly deflected by the strong magnetic field at the edge of the island (or equivalently, the X-point) and is then trapped in the island. In contrast, the blue particle is basically unaffected by the magnetic field.  The oscillations prior to entering the island are Speiser orbit trajectories characteristic of particle motion in X-points. Note that before entering the island, the largest component of the particles' momentum is in the $\pm z$ direction (not shown) due to acceleration by the electric field. \label{fig:structureschema}}
	
\end{figure}

 We approximately estimate that deflection by the strong magnetic field at the edge of an island (or equivalently, X-point) will occur if  the particle is strongly affected by the magnetic field $B_{\rm c}$ at the edge of a CSS before it can escape from the structure.  The time required for escape is just $D_{\rm c}/\langle v_y\rangle$, where $D_{\rm c}$ is the length of the current sheet structure in the $y$ direction  and $\langle v_y\rangle$ is the average velocity of the particle in the $y$ direction as it crosses the structure. To estimate $\langle v_y\rangle$, we calculate the average velocity $\langle v_y\rangle_{\rm r}/c$ for our test particles  in Simulation {\tt S30} with $\sigma=30$ during their rapid acceleration phases in X-points. We show in Table 2 that $\langle v_y\rangle_{\rm r}/c\approx 0.50$ on average before saturation and $\langle v_y\rangle_{\rm r}/c\approx 0.72$ afterwards, which is already quite large. Particles passing through a region of high magnetic field at the edge of a CSS will have an even larger velocity in that direction because the magnitude of $v_y$ increases monotonically during the rapid acceleration phase. Therefore, we set $\langle v_y\rangle=c$ in this part of the analysis.

We estimate the time required for the particle to be deflected as the time for the direction of the particle's momentum to change by an angle of $\pi/2$. When the particle is deflected that much, its momentum in the $y$ direction is close to 0 (because the particle's momentum was originally mostly in the $y$ direction) and it is no longer able to escape. This time is 

\begin{equation}\frac{\gamma m c } {4 q B_{\rm c}}=\frac{\gamma}{4} \frac{B_0}{B_{\rm c}}\frac{r_{\rm L}}{c},\end{equation}
where we have used the definition of $r_{\rm L}$ in the second part of the expression.

We can now parameterize each CSS using the characteristic Lorentz factor $\gamma_{\rm c}$ for which the escape and deflection times are equal, given by

\begin{equation}
\gamma_{\rm c}=4\sigma \frac{D_{\rm c}}{ \sigma  r_{\rm L}}\frac{B_{\rm c}}{B_0}. \label{eq:threshold}
\end{equation}
This equation shows that larger CSS with higher magnetic field also have higher characteristic Lorentz factors. 

We schematically show in Figure  \ref{fig:structureschema} how the deflection process works in the case of a magnetic island. The blue particle with  $\gamma \gg \gamma_{\rm c}$ is unaffected by the magnetic island and can undergo further acceleration in another X-point. In contrast, the red particle with $\gamma \ll \gamma_{\rm c}$ is strongly deflected and then trapped in the island. As a result, it can no longer undergo significant acceleration.  The comparison shows that particles only "see" current sheet structures with $\gamma_{\rm c}>\gamma$. We show in Section
\ref{sec:implicitcurcalc} how this affects the saturation of particle acceleration.

\subsubsection{Current sheet analysis}

\begin{figure}
	\begin{center}
		\includegraphics[width = 0.48\textwidth]{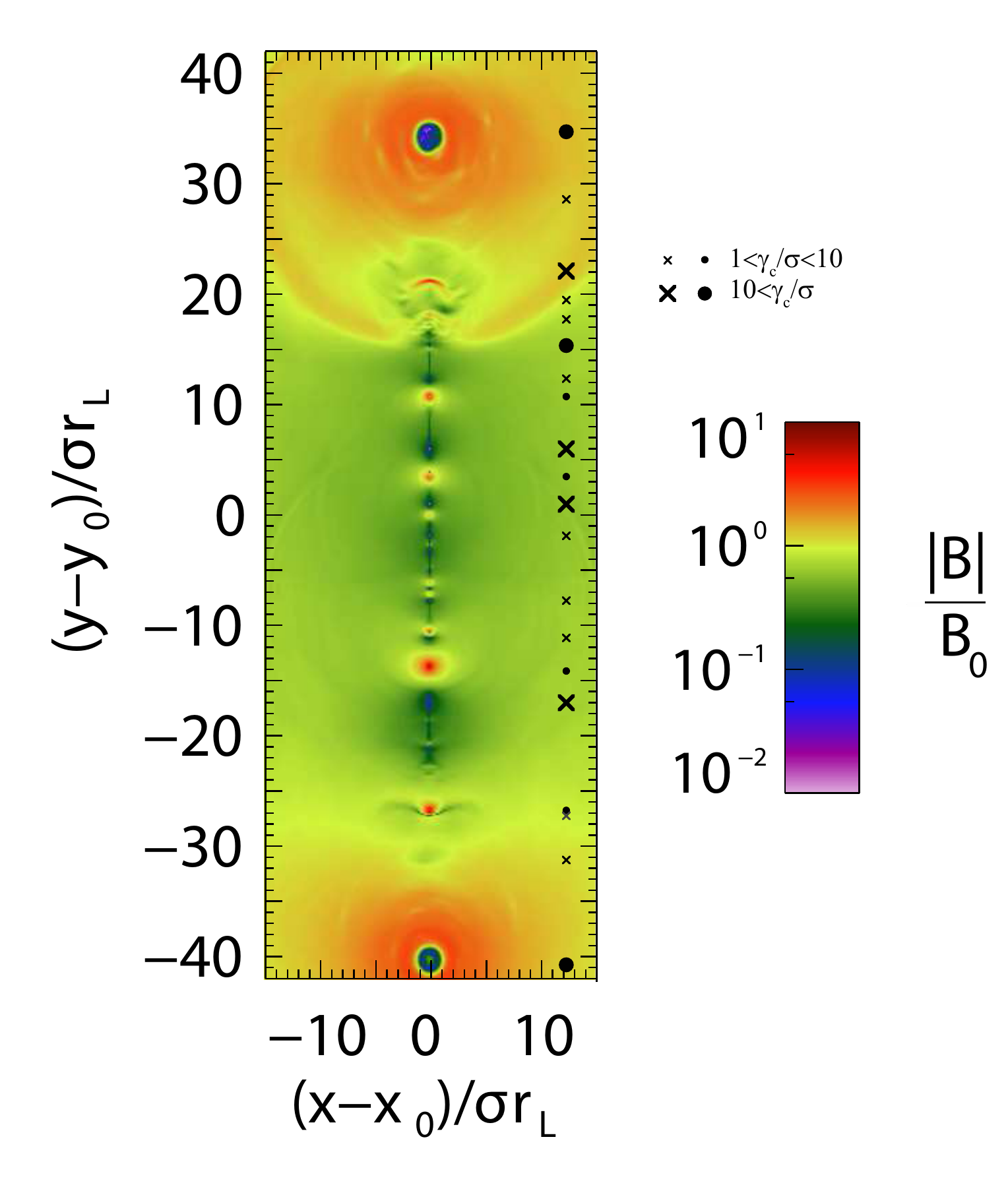}
	\end{center}
	\caption{ The identification of current sheet structures using our algorithm. The colors show the local magnetic field normalized to the background field $B_0$  in the Simulation {\tt S30} at the time $\tau\approx140$ of post-saturation tracing. We focus on a region centered at the middle location of particle injection $(x_0,y_0)=(240,110.75) \sigma r_{\rm L}$ at that time.   The locations $y_{\rm min}$ of current sheet structures are shown to the right of the main figure. The size of these symbols corresponds to which of two ranges of $\gamma_{\rm c}$ (shown to the right of the figure) the CSS corresponds to. X-points are labeled with an X and magnetic islands with a filled circle. They are differentiated by finding whether the structure corresponds to a maximum  (island) or a minimum (X-point) in number density as well, although this identification is uncertain near the edges of the primary X-point at $|y-y_0|>17\sigma r_{\rm L}$.   \label{fig:cursheetid}}
	
\end{figure}

We can now investigate the structure of the current sheet to find $D_{\rm c}$ and $B_{\rm c}$ for each structure. We first find  the maxima and minima of the reconnected  field $|B_{x}|$ over the length of the current sheet. Each of the minima corresponds to the center of a CSS, while maxima correspond to the edges of a magnetic island or X-point.  We hierarchically pair the maxima and minima in order to associate each location of large deflection with a CSS. For each pair we can calculate $D_{\rm c}=2(y_{\rm max}-y_{\rm min})$  and $B_{\rm c}=(B_{\rm x, max}-B_{\rm x,min})/2$.  The factor of 2 in the expression for $D_{\rm c}$ arises because the size of the structure is approximately the distance between two maxima (e.g., from one side of a magnetic island to another) or two minima (from an X-point to an island), not that between a maximum and a minimum. The factor of 1/2 in the expression for $B_{\rm c}$ arises because we estimate that the average magnetic field in the structure is around half of the maximum field.  Finally, we can calculate $\gamma_{\rm c}$ using $D_{\rm c}$, $B_{\rm c}$, and Equation (\ref{eq:threshold}).

Figure \ref{fig:cursheetid} shows the locations $y_{\rm min}$ for highly significant CSS with $\gamma_{\rm c}>\sigma$ at the beginning of post-saturation particle tracing for Simulation {\tt S30}.   The minima calculated using our algorithm clearly correspond to minima of $B_0$. These CSS typically be identified as X-points or magnetic islands from inspection or by reference to the simulation's density structure (as discussed in the figure caption). Note, however, that this identification has no effect on our results, which are based on generic properties of CSS.

\begin{figure}
\begin{center}
\includegraphics[width = 0.47\textwidth]{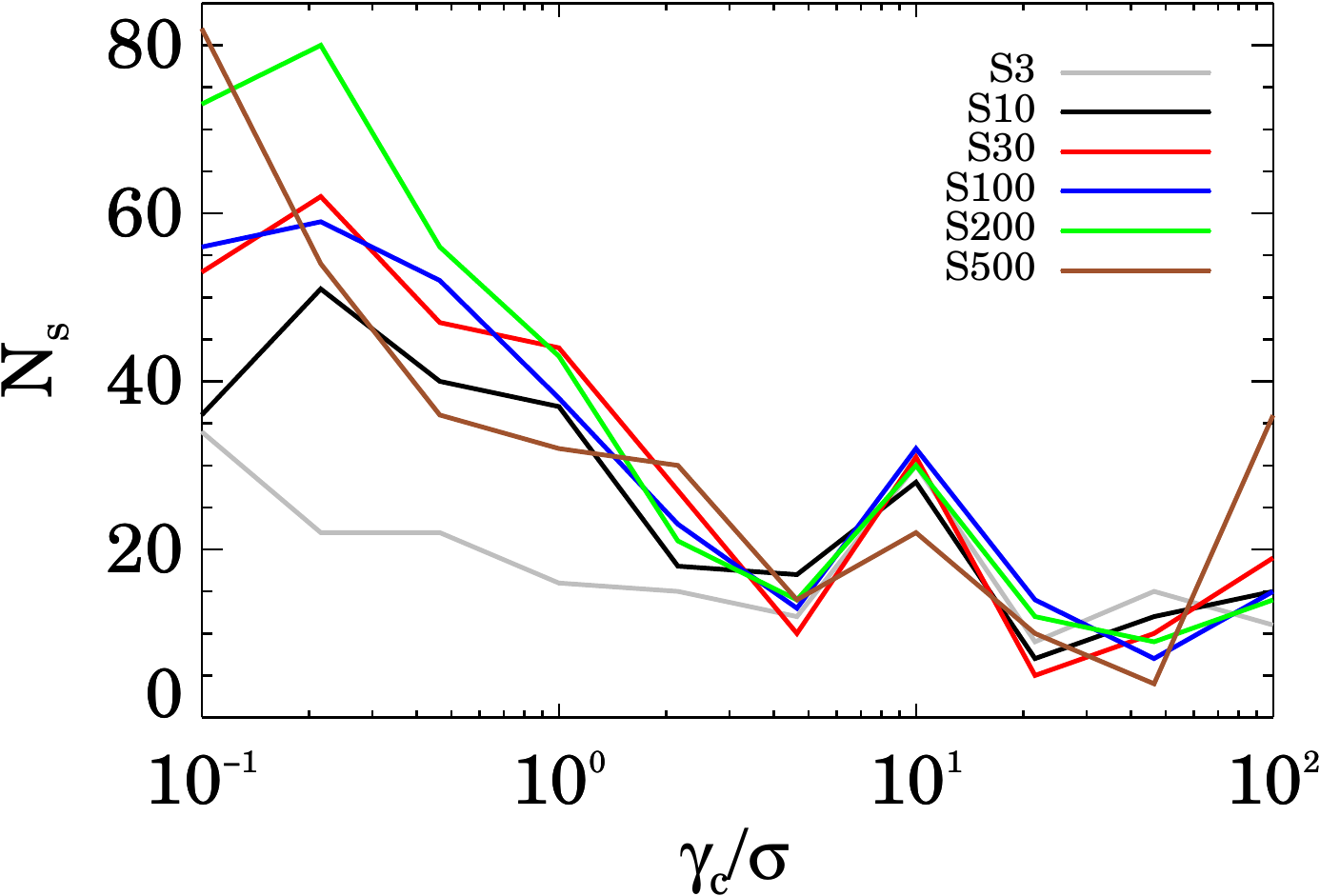}
\end{center}

\caption{ The number of CSS $N_{\rm s}$ at each value of $\gamma_{\rm c}/\sigma  $ summed over 3 outputs near the time of saturation $\tau_{\rm s}\approx84$. We multiply the number of structures by 2 for Simulation {\tt S500}, because the normalized system size in that simulation was half of that found in the other simulations. It is clear that the distribution does not depend strongly on $\sigma$ for $\sigma>3$. \label{fig:gamchist}}
\end{figure}

Figure \ref{fig:gamchist} shows  $N_{\rm s}$, the number of CSS with each value of $\gamma_{\rm c}$ in all of the simulations  at the time of saturation $\tau_{\rm s}\approx84$. It is clear that the distribution of $\gamma_{\rm c}/\sigma$ does not depend significantly on $\sigma$ for $\sigma>3$. There are many small CSS with $\gamma_{\rm c}/\sigma\sim 0.1$, but the number of such structures quickly declines with increasing $\gamma_{\rm c}$, leveling out at approximately $\gamma_{\rm c}=4 \sigma$. Because increased particle acceleration no longer reduces the number of disruptive CSS encountered beyond this point, our result suggests that saturation of particle acceleration may occur near $\gamma\approx4\sigma$. We return to this argument in Section \ref{sec:implicitcurcalc}.

\subsubsection{Current sheet structures and the saturation of particle acceleration}\label{sec:implicitcurcalc}

 We now consider the maximum energy that can be reached by a particle accelerated in a CSS. Assuming that $\mathbf{E} \cdot \mathbf{v} \approx E c$ is approximately constant over a typical particle's trajectory, the typical Lorentz factor $\gamma_{\rm t}$  reached by a  particle accelerated in single X-point is approximately 
 \begin{equation}
\gamma_{\rm t} = \frac{q E}{m c}\frac{{D}_{\rm c}}{v_y}.
\end{equation}

This would give the correct final Lorentz factor if particles were accelerated in only one structure. But particles do not finish their acceleration unless they encounter a CSS with $\gamma_{\rm c}>\gamma$. Thus, the size of the acceleration region actually encountered by a particle depends on the particle energy.  The typical size of acceleration regions for particles with $\gamma\sim \gamma_{\rm c}$ can be estimated as $\langle D\rangle(\gamma_{\rm c})$, defined as the median distance between CSS with structure parameter larger than $\gamma_{\rm c}$. We normalize this quantity to $\sigma r_{\rm L}$ in our calculations, as we do with all length scales in this paper.

\begin{figure}
	\begin{center}
		\includegraphics[width = 0.48\textwidth]{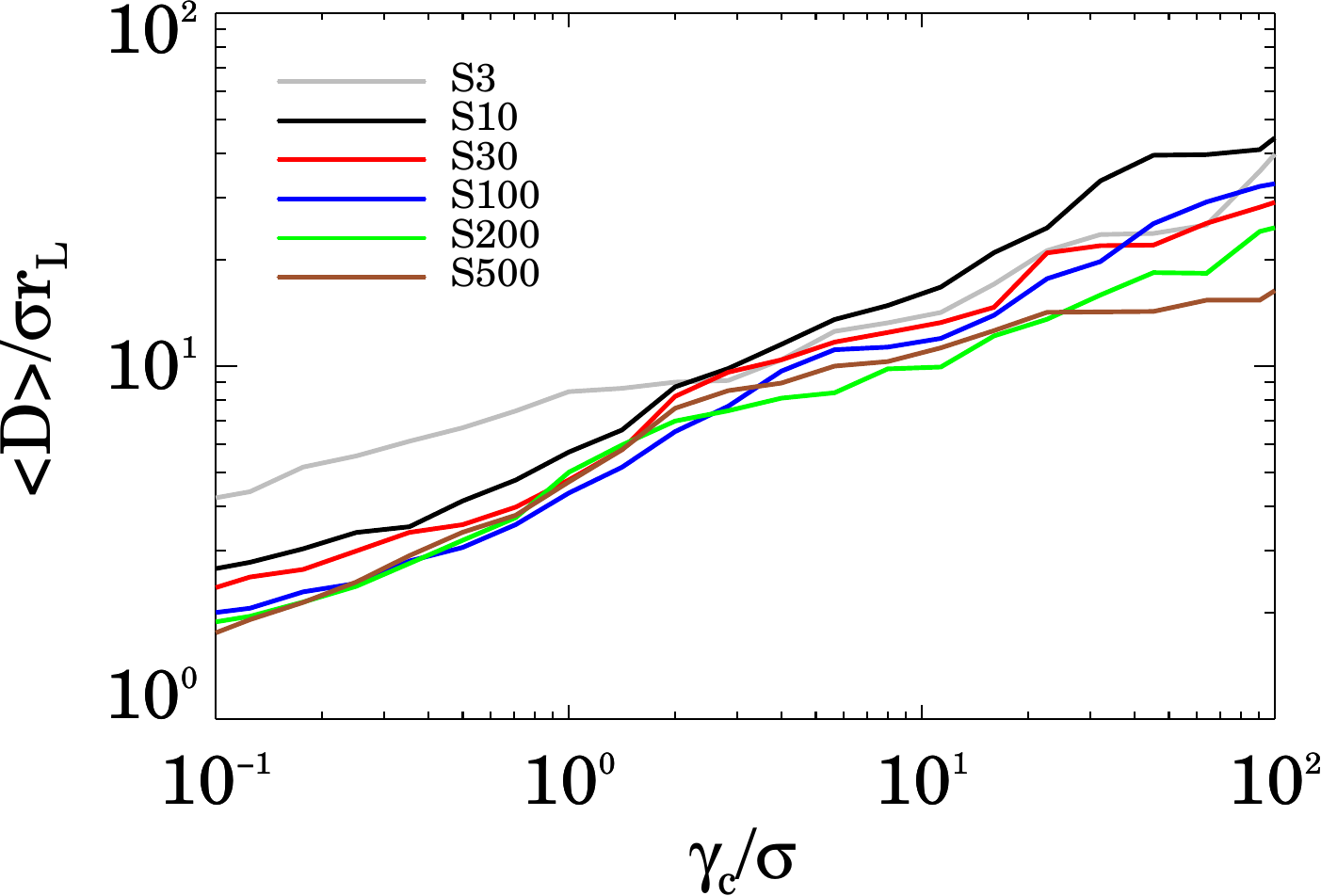}
	\end{center}
	\caption{ The median distance $\langle D\rangle/\sigma r_{\rm L}$ between CSS with parameter larger than $\gamma_{\rm c}$ for all simulations. This distance is clearly independent of $\sigma$. For low $\gamma_{\rm c}$ this distance grows  as $\gamma_{\rm c}^{1/3}$, but it levels out  at higher $\gamma_{\rm c}$.  \label{fig:dbarout}}
\end{figure}
Figure \ref{fig:dbarout} shows how this parameter varies with $\gamma_{\rm c}$. It is clear that the distance between CSS increases slowly and monotonically with $\gamma_{c}$ approximately as $\gamma_{\rm c}^{1/3}$ at low $\gamma_{\rm c}$, leveling off somewhat at higher $\gamma_{\rm c}$.  $\langle D\rangle/\sigma r_{\rm L}$  is basically independent of $\sigma$ for $\sigma>3$, just as $N_{\rm s}$ is.  We note that $10\sigma r_{\rm L}$ is the approximate size of CSS predicted by \citet{werner_16} based on a heuristic argument using the results of \citet{larrabee_03} and \citet{kirk_03}. Here, we confirm that this result is approximately correct for $\gamma_{\rm c}\approx 4 \sigma$ and somewhat low for $\gamma_{\rm c}>10 \sigma$.

Using this parameter, we find an implicit equation for the maximum Lorentz factor $\gamma_{\rm max}$ that can be reached by particles in CSS with parameter $\gamma_{\rm c}$ or greater:
 \begin{equation}
 \gamma_{\rm max}(\gamma_{\rm c}) = \frac{q E}{m c}\frac{\langle D\rangle(\gamma_{\rm c})}{\langle v_y\rangle}.
\end{equation}

Rearranging this equation using the definition of $r_{\rm L}$ yields

\begin{figure}
	\begin{center}
		\includegraphics[width = 0.48\textwidth]{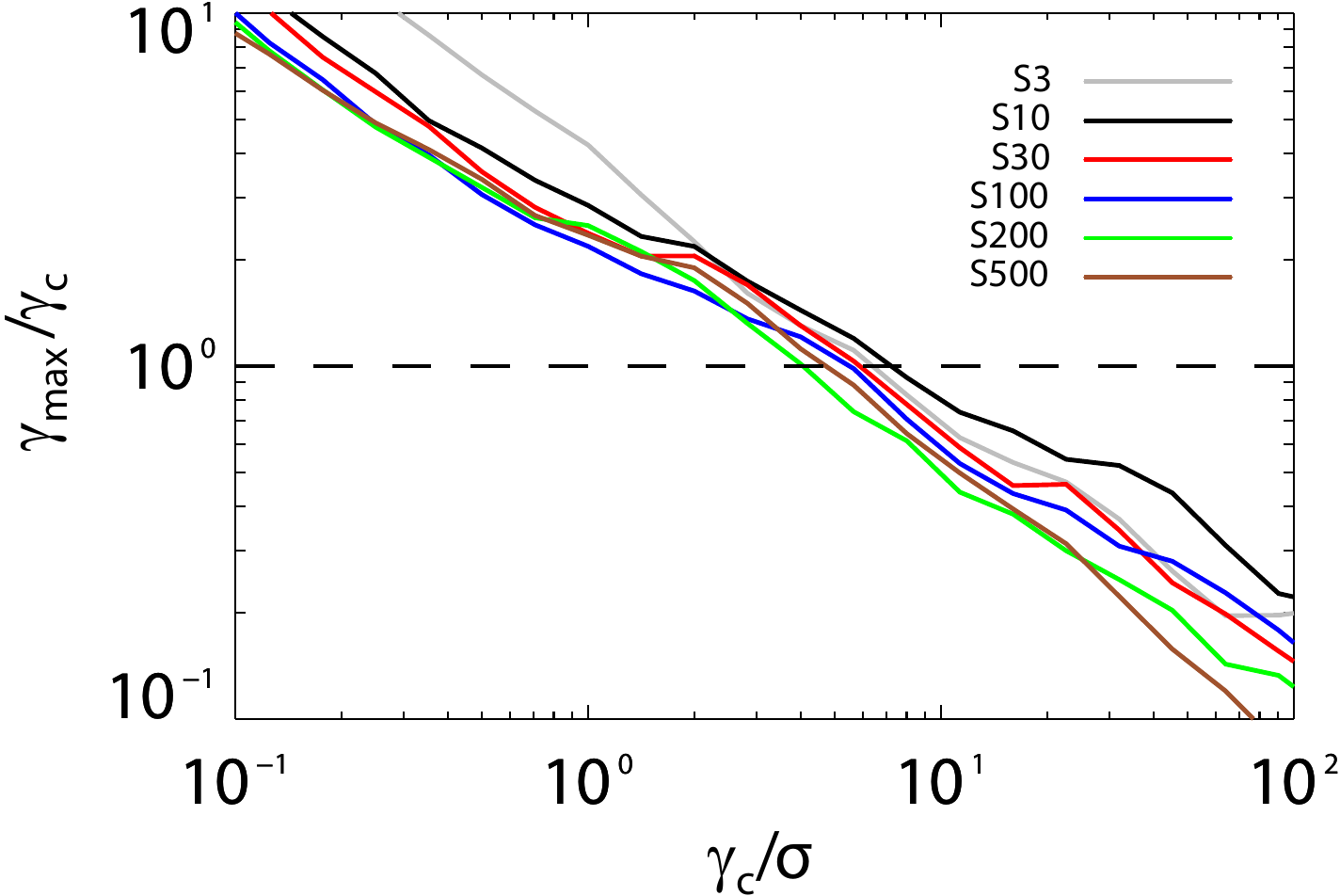}
	\end{center}
	\caption{ The ratio $\gamma_{\rm max}/\gamma_{\rm c}$ as a function of $\gamma_{\rm c}$.  When the ratio falls below 1 (shown with the dashed line), particle acceleration above $\gamma$ is no longer possible  due to the intervention of significant CSS.\label{fig:gamtout}}
\end{figure}
\begin{equation}
\frac{\gamma_{\rm max}(\gamma_{\rm c})}{\sigma} = \frac{c E}{\langle v_y \rangle B_0}\frac{\langle D\rangle(\gamma_{\rm c})}{ \sigma  r_{\rm L}}.\label{eq:gammafpred}
\end{equation}

We assume that $E/B_0$ is not significantly dependent on $\sigma$ because reconnection rate has no dependence on $\sigma$ \citep{kagan_16}. We also assume that $c/\langle v_y \rangle$ does not depend significantly on $\sigma$ because the bulk outflow momentum from reconnection regions is independent of $\sigma$ \citep{melzani_14, kagan_16}. We use parameters from our particle tracing in Simulation {\tt S30}  (Table 2) during the rapid phase of acceleration to estimate these values, choosing $E/B_0=0.25$ and $\langle v_y\rangle/c=0.5$ based on the pre-saturation averages. Note that the product $c E/\langle v_y \rangle B_0$ is very similar if we instead use the post-saturation averages. For our chosen parameters, Equation (\ref{eq:gammafpred})  reduces to 
\begin{equation}
\frac{\gamma_{\rm max}(\gamma_{\rm c})}{\sigma} = 0.5\frac{\langle D\rangle(\gamma_{\rm c})}{ \sigma  r_{\rm L}}.\label{eq:gammafsimp}
\end{equation}

 Because $\langle D\rangle/\sigma r_{\rm L}$ is independent of $\sigma$ as seen in Figure \ref{fig:dbarout}, this equation indicates that $\gamma_{\rm max}\propto \sigma$. Thus, the spontaneous formation of CSS with a separation proportional to  $\sigma  r_{\rm L}$ provides a physical explanation for the saturation of the power law in our fits.

While Equation (\ref{eq:gammafsimp}) is still implicit, it is clear that there are two regimes for $\gamma_{\rm max}$ depending on the ratio $\gamma_{\rm max}/\gamma_{\rm c}$. If  $\gamma_{\rm max}>\gamma_{\rm c}$, particle acceleration in X-points with parameter of at least $\gamma_{\rm c}$ tends to result in the particle reaching high enough energy to be unaffected by a significant number of large CSS. This makes further acceleration more probable. In contrast, if $\gamma_{\rm max}<\gamma_{\rm c}$ particles will not accelerate enough to reduce their susceptibility to deflection by CSS. As a result, we expect that such particles will not accelerate beyond $\gamma_{\rm max}$. This will produce a cutoff at $\gamma_{\rm max}=\gamma_{\rm c}$.

In Figure \ref{fig:gamtout}, we plot the ratio $\gamma_{\rm max}/\gamma_{\rm c}$ as a function of $\gamma_{\rm c}$. The figure shows that $\gamma_{\rm max}/\gamma_{\rm c}$ decreases monotonically with $\gamma_{\rm c}$ approximately as $\gamma_{\rm c}^{-2/3}$. Figure \ref{fig:gamtout} shows that $\gamma_{\rm max}=\gamma_{\rm c}$ at around $\gamma_{\rm c}\approx5\sigma$, which is close to the location of the cutoff that we find in fits of the power law particle energy spectrum. 

Figure \ref{fig:dbarout} shows that value of $\gamma_{\rm max}=\langle D\rangle/2\sigma r_{\rm L}$ is typically around $30\sigma$ for $\gamma_{\rm c}=100 \sigma$. Because particles accelerating in large X-points with $\gamma_{\rm c} \gg 4 \sigma$ cannot reach higher Lorentz factors than $\gamma_{\rm max}$, we expect almost no particles to reach above $\gamma=30 \sigma$, which is less than a factor of 10 higher than the cutoff. This indicates that the cutoff should be sharp, consistent with an exponential or super-exponential.   Thus, our current sheet analysis has correctly predicted the existence and the location at $\gamma\approx 4-5 \sigma$ of  a sharp high-energy cutoff of the power law in the particle energy spectrum produced in relativistic reconnection.

It has commonly been thought that because current sheet evolution produces larger and larger primary X-points and magnetic islands, the properties of the particle energy spectrum in reconnection are determined by global dynamics. But we have shown that spontaneous tearing in large X-points continually produces magnetic islands with a characteristic size of $\sim  \sigma r_{\rm L}$, which implies that the maximum $\gamma\sim\sigma$. Our detailed analysis gives a more precise cutoff at $\gamma\approx 4 \sigma$. Thus, the small-scale properties of the current sheet can never be ignored in investigating reconnection even though the global dynamics dominate the overall structure. Constraints from this small-scale structure have significant effects on reconnection models of astrophysical emission, as we show in the next section.

\section{Application to astrophysical sources}

\label{sec:astrophysobj}
We now discuss the implications of our results in Section 3.1 for the application of relativistic reconnection to observed radiation from high-energy astrophysical sources. We note that these conclusions may be unreliable if new acceleration mechanisms become important in very large scale systems. In addition, the application of our results to ion-electron plasmas still needs to be confirmed.  In this section, we assume that our conclusions scale to very large scales found in astrophysical systems and that our pair-plasma results apply to any ion-electron plasmas that may be present in such systems..

 We have shown in Section 3.1.1 that the power law index found in reconnection falls in the range $1.15<p<2.3$. The power law index of synchrotron radiation resulting from particles in the power law is typically either $-p/2$ (if cooling is fast) or  $-(p-1)/2$ (if cooling is slow) \citep{sari_spectra_1998}.  This corresponds to a range of flux per logarithmic interval $\nu F_{\nu}\propto\nu^{-0.15}-\nu^{0.9}$. Thus, reconnection produces nearly flat or rising logarithmic radiation spectra.  In Section 3.1.2, we have shown that the extent of the power law in the particle energy spectrum is quite narrow:  $\gamma_f/\gamma_i\le40$. The corresponding maximum dynamic range of the high-energy synchrotron spectrum emitted by the particles is $\approx40^2=1600$.

This narrow range can be expanded  if significant relativistic bulk flows or variations in the magnetic field are present because the energy at the synchrotron peak is proportional to $B \Gamma$. Ultrarelativistic bulk flows with $\Gamma \gg 1$ are rarely produced in relativistic reconnection \citep{melzani_14,guo_14,kagan_16}, although they may occur for some initial conditions \citep{sironi_16}. The variation in $B$ in the outskirts of magnetic islands where particles emit is also  relatively small, typically no more than a factor of a few (as can be seen in Figure 7). Overall, we estimate that $B \Gamma$ does not vary by more than a factor of 5 for particles in a given reconnection region.  Variability in the central source or resulting from global instabilities may also be a source of variation in $B\Gamma$. If we roughly estimate that the variation in $B \Gamma$ from these sources is similar to the variation in the location of the spectral peak, we can estimate the size of this effect for any given source.

We now consider whether these properties of reconnection are consistent with various observed astrophysical systems. We first consider the flares in the Crab's PWN \citep{tavani_11,abdo_11,striani_11,buehler_12,mayer_13}, which extend for around 1.5 decades in frequency below the peak of the distribution, with a sharp cutoff above it. It is possible that the power law extends a bit further at low energies where it is swamped by the quiescent spectrum, but this dynamic range is still easily consistent with reconnection. The flare spectra near the peak frequency  are nearly flat in $\nu F_{\nu}$ at the peak time of the flares. Therefore, the reconnection model with a relatively low $\sigma$ works well for the Crab flares. 

The TeV flares in AGN \citep{aharonian_07,aharonian_09,albert_07,tavecchio_13, cologna_17} have a power law spectrum with some curvature that extends for around 2 decades in frequency. This range can easily be produced by reconnection.   The intrinsic power law spectra of Fermi-LAT blazars have hard power law indices in the range $\nu F_{\nu}\propto \nu^{-0.65}-\nu^{-0.15}$\citep{singal_flux_2012}. Flaring galaxies have quiescent spectra consistent with these values, but the flaring spectra are typically harder \citep{abramowsky_12,albert_08, cologna_17}. Thus, most TeV flares are consistent with being produced in reconnection at low $\sigma$, or even at moderate $\sigma$ in extraordinary cases.

Finally, we consider the prompt phase of gamma-ray bursts.  In many GRBs in the Fermi catalog, the combined GBM and LAT data are consistent with a single power law component over a large frequency range. In GRB 0901003, the prompt emission observed by Fermi appears to constitute a single power law from a peak at $\sim 400$ keV in the GBM instrument all the way up to $\sim $2.8 GeV as observed in the LAT instrument \citep{zhang_fermi11}.   The peak of the energy spectrum varied by a factor of only 1.5 during the burst, so variation in  $B \Gamma$ likely cannot expand the power law range of reconnection (which is less than 1600) enough to explain the observed dynamic range of 7000 for the GRB.  While this GRB is an extreme case, many other GRBs observed in the LAT band  have a directly observed dynamic range of around 1000-3000 in frequency with no strong evidence of a cutoff beyond that range \citep{Ackermann_13}. It not easy to explain such GRBs in a reconnection model.

 Additional evidence against the direct application of reconnection to prompt GRB emission comes from spectral slopes. Fits to the high-energy portion of the spectrum of gamma-ray burst range from $\nu F_{\nu}\propto \nu^{-2.8}-\nu^{-0.6}$ \citep{gruber_14}, significantly softer than in reconnection even at low $\sigma$.  Thus, it seems difficult for relativistic reconnection to produce the nonthermal emission seen in the prompt phase of GRBs. In summary, it appears that relativistic reconnection is clearly applicable to the Crab flares and to TeV flares in AGN, and has difficulty accounting for the prompt GRB emission.
\section{Conclusions}

\label{sec:conclusions}
In this paper, we investigated the saturation of the high-energy power law spectrum in reconnection using particle-in-cell simulations at various magnetizations $\sigma$. We used particle tracing to identify the dominant particle acceleration  
mechanism as direct acceleration  by the reconnection electric field in X-points. We then analyzed the structure of the current sheets in our simulations  to show that the spontaneous formation of secondary islands and X-points (referred to as CSS, current sheet structures) was responsible for saturation.

Our conclusions are as follows:

\begin{itemize}

\item The high-energy part of the particle energy spectrum produced by reconnection can be fit with a hard power law followed by a super-exponential cutoff.

\item The high-energy power law in the particle particle energy spectrum saturates at $\gamma_f\approx 4\sigma$  This saturation occurs at  the normalized time $\tau\approx84$, consistent with the saturation at $L=40 \sigma r_{\rm L}$ found by \citet{werner_16}. 

\item The ratio $\gamma_f/\gamma_i$, where $\gamma_i$ is the minimum energy of the power law, approaches $\approx40$ at large $\sigma$.

\item Our particle tracing reveals that X-points are responsible for the majority of particle acceleration, especially in a logarithmic sense. The average energy of particles after a single episode of rapid acceleration in the X-point is at least $70\%$ of the average energy of all accelerated particles after saturation.

\item We find that the saturation of particle acceleration in reconnection is due to the spontaneous production of magnetic islands within large X-points. At the edges of the magnetic islands, strong magnetic fields are present that can deflect particles and end particle acceleration. We characterize current sheet structures (CSS) including both X-points and magnetic islands in a unified way by the maximum Lorentz factor $\gamma_{\rm c}$ of particles that the adjacent strong magnetic fields can deflect. 

\item The distribution of $\gamma_{\rm c}/\sigma$ is largely independent of $\sigma$. It declines quickly at small $\gamma_{\rm c}$  but plateaus at $\gamma_{\rm c}\approx4 \sigma$. This is consistent with saturation of the power law at $\gamma_{\rm f}\approx 4 \sigma$.

\item We calculate an implicit equation for the maximum acceleration possible in CSS as a function of their typical size and $\gamma_{\rm c}$ parameter.  We find that particles entering the current sheet can be accelerated to a maximum energy of approximately $5\sigma$ before they are deflected by an encounter with a significant CSS and their acceleration is stopped. This is very close to the location of the high energy cutoff we find in our fits.

\item Our results indicate that the fundamental spatial scale for particle acceleration is of order $\sigma r_{\rm L}$. Because secondary islands and X-points are constantly produced at this scale, it remains important even at late times when primary islands and X-points are much larger than  $\sigma r_{\rm L}$, 

\item Our simulations predict that particles accelerated in reconnection will produce synchrotron spectra that are hard power laws with index $\nu F_{\nu}\approx \nu^0$ and a narrow frequency range of $\sim 1600$.  TeV AGN flare spectra and Crab PWN flare spectra, which are hard and have narrow frequency ranges, can easily be produced by reconnection. In contrast, the  relatively soft and broad power laws present in prompt GRB emission are difficult to produce in reconnection.

\end{itemize}

  \section*{Acknowledgements}

 This research was supported by the I-CORE Center for Excellence in Research in Astrophysics and by an ISA grant. DK and TP were partially supported by a CNSF-ISF grant. EN was partially supported by an ISF grant and by an ERC starting grant.

\bibliographystyle{mnras}

\bibliography{saturation}

\end{document}